\documentclass[twocolumn]{revtex4-2}

\usepackage{hyperref,lineno,graphicx}

\def\Authors{Jorge P. Rodr\'iguez \,$^{1,2*}$, Alberto Aleta\,$^{2,3}$ and Yamir Moreno\,$^{2,3,4}$}

\begin{document}
 
\title{Digital cities and the spread of COVID-19: characterizing the impact of non-pharmaceutical interventions in five cities in Spain}
\author{\Authors} 
\address{$^{1}$Instituto Mediterr\'aneo de Estudios Avanzados (IMEDEA), CSIC-UIB, Esporles, Spain \\
$^{2}$Institute for Biocomputation and Physics of Complex Systems, University of Zaragoza, Zaragoza, Spain \\
$^{3}$Department of Theoretical Physics, University of Zaragoza, Zaragoza, Spain\\
$^{4}$CENTAI Institute, Turin, Italy}

\begin{abstract}
Mathematical modeling has been fundamental to achieving near real-time accurate forecasts of the spread of COVID-19. Similarly, the design of non-pharmaceutical interventions has played a key role in the application of policies to contain the spread. However, there is less work done regarding quantitative approaches to characterize the impact of each intervention, which can greatly vary depending on the culture, region, and specific circumstances of the population under consideration. In this work, we develop a high-resolution, data-driven agent-based model of the spread of COVID-19 among the population in five Spanish cities. These populations synthesize multiple data sources that summarize the main interaction environments leading to potential contacts. We simulate the spreading of COVID-19 in these cities and study the effect of several non-pharmaceutical interventions. We illustrate the potential of our approach through a case study and derive the impact of the most relevant interventions through scenarios where they are suppressed. Our framework constitutes a first tool to simulate different intervention scenarios for decision-making.   

\end{abstract}
\maketitle

\section{Introduction}
The COVID-19 pandemic has globally impacted a plethora of systems, with health \cite{barber2022estimating}, socio-economic \cite{bonaccorsi2020economic,pak2020economic}, and environmental \cite{bates2021global} consequences. To control the spread of SARS-CoV-2, policymakers implemented a diversity of procedures, grouped into either mitigation or suppression strategies. Lockdowns, implying home confinement, were frequently introduced to stop the spreading in early 2020 when the dynamics of the infection mechanisms was not clear. However, these lockdowns resulted in deep impacts on the economy, and later on, other non-pharmaceutical interventions were designed, such as the use of face masks, the closure of restaurants, universities, or schools, as well as contact tracing, testing, and isolation of close contacts of infected individuals. 

The initial stages of the pandemic represented a high degree of uncertainty, both regarding the original transmission of the pathogen to human beings and reliable surveillance data (due to low testing efforts and inappropriate surveillance systems \cite{Starnini2021}). Nowadays the situation has improved, as the availability of more data - even if many times of poor quality and low reliability- in principle allows to characterize the spreading at a large scale. Moreover, the existent data enables the development of mathematical models that help quantify the observed evolution of the pandemic and evaluate the effects of the intervention scenarios. 

The first wave of COVID-19 represented a challenge for modeling approaches due to the general bad data quality. Specifically, the lack of knowledge about the COVID-19 spread, the similarity between the symptoms of COVID-19 and those of influenza, and the low testing effort led together to lower rates of diagnosis and hence underreporting mainly in the number of cases, but also in the number of deaths. Seroprevalence studies \cite{pollan2020prevalence} and the analysis of anomalies on the temporal series of deaths \cite{garcia2021retrospective} were needed to estimate the real impact of the spreading process, showing that there were up to 10 times more cases than the reported ones. In this regard, spreading models can shed light on the real outcome of the infection across the population.

To properly model the spreading of a disease in the population, it is fundamental to acknowledge that human interactions are highly heterogeneous. Although network epidemiology can capture part of this diversity, such as the broad nature of the distribution of the number of interactions, the variability of contexts remains out of this formalism. These contexts can be effectively captured using multilayer networks, which are networks with multiple layers, each one describing the interactions in a different context \cite{kivela2014multilayer,aleta2019multilayer}. In this work, we leverage anonymous, publicly available data to build high-resolution synthetic cities and encode them in multilayer networks \cite{Fumanelli2012Sep,Mistry2021Jan}. We use these synthetic networks to study the propagation of the first wave of COVID-19 in five Spanish cities. Furthermore, we extend the simulation to the second wave for the particular case of the city of Zaragoza and thoroughly characterize the impact of non-pharmaceutical interventions during this period.

\section{Materials and methods}
\begin{figure*}
	\includegraphics[width=\textwidth]{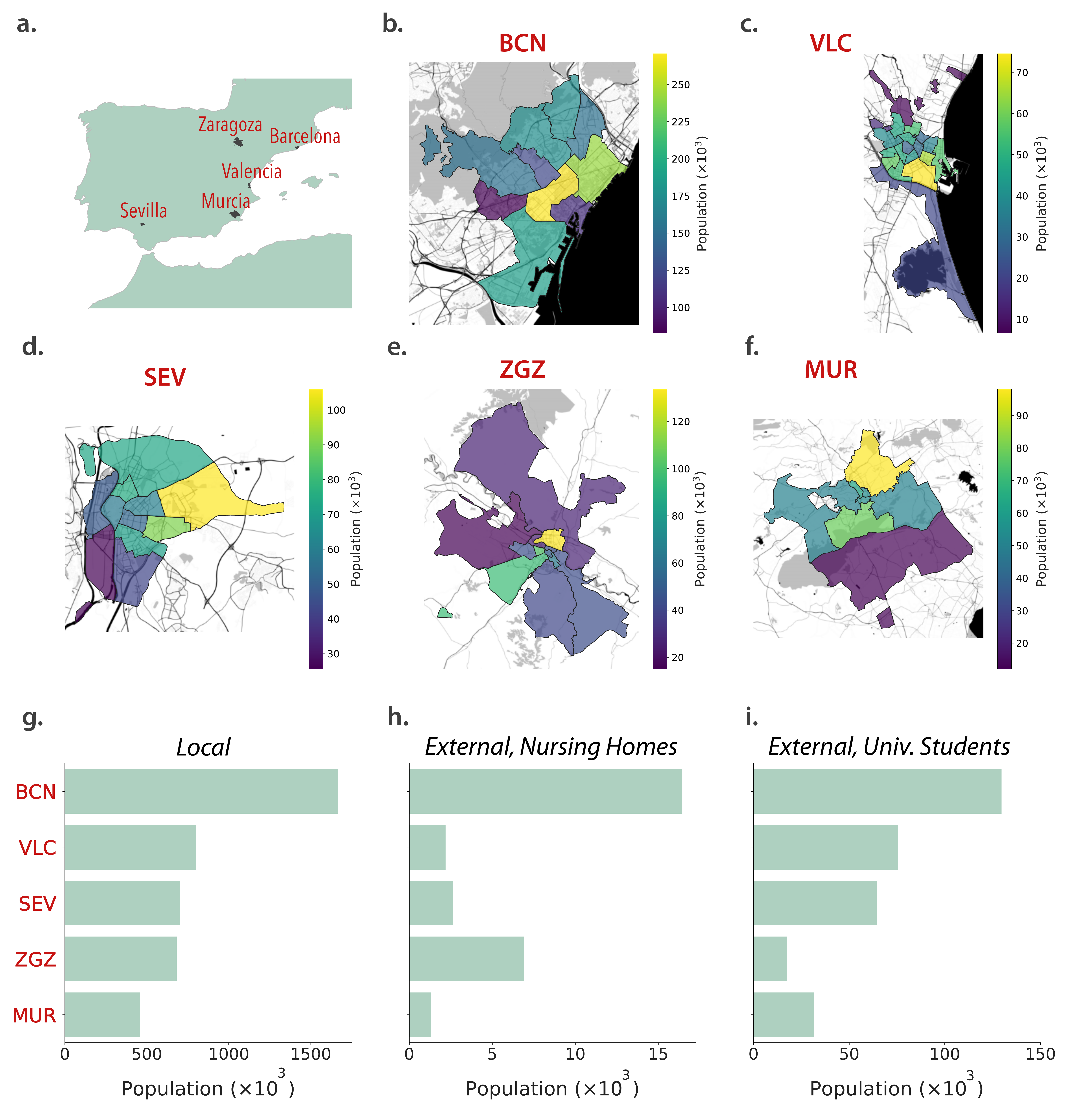}
	\caption{Geography and demography of the five cities represented through synthetic populations. {\bf a,} Location of the five cities. {\bf b-f} Geographical representation (Map tiles by Stamen Design, under CC BY 3.0. Data by OpenStreetMap, under ODbL) of the district structure at each city (source: Instituto Nacional de Estad\'istica, INE), with colors representing the population of each district. {\bf g} Local and external populations (nursing homes and university students). The five cities studied are Barcelona (BCN), Valencia (VLC), Seville (SEV), Zaragoza (ZGZ), and Murcia (MUR).}
	\label{fig1}
\end{figure*}

We create five digital populations describing the interactions between the inhabitants in the cities of Barcelona, Valencia, Seville, Zaragoza, and Murcia, all of them located in Spain (Fig. \ref{fig1}a). Their population ranges between 450 thousand and 1.7 million inhabitants. Additionally, we include external individuals that may not be registered in the census but with most of their interactions expected to happen in these cities. These external individuals include old people living in nursing homes and non-local university students. The specific data sources for each city and each feature are listed in the Supplementary Material.
  
{\bf Demography. } We obtained the geographical distribution, sex, and age of the inhabitants of the cities at the beginning of 2020 from multiple demographic data sources. The maximum spatial resolution was the census district (Fig. \ref{fig1}b-f), at which we found most of the needed information to create the synthetic digital cities. Ages were available in age groups with a resolution of 5 years. Thus, we interpolated these age groups to consider a resolution of 1 year between 0 and 30 years, which was necessary to properly infer the interactions at schools and universities. This allowed us to create a synthetic population for each city resembling the characteristics of the real ones.

\begin{figure*}
	\centering
	\includegraphics[width=\textwidth]{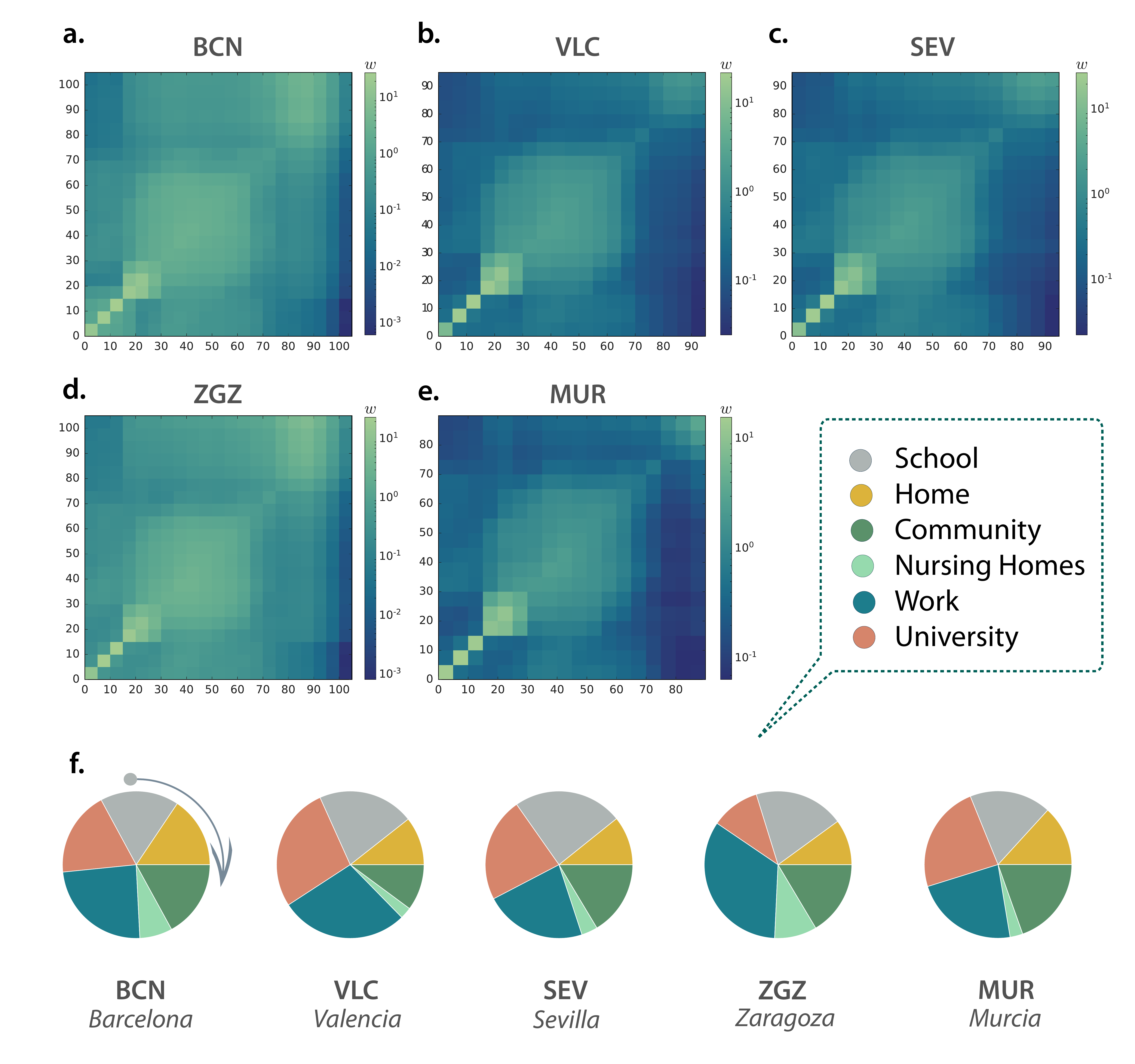}
	\caption{Contact matrices between age groups in each of the cities considered in this study. a-e, Each entry is computed including the total number of observed links between the row and the column age groups, normalized by the number of people in each column, such that they represent, for a given column, the expected number of links of a random individual to individuals from each row. f, Link distribution among the different layers in the five studied cities. }
	\label{fig2}
\end{figure*}

{\bf Contact networks.} We modeled the contacts between individuals through networks described by the aggregation of multiple interaction layers \cite{aleta2020modelling}. Specifically, we considered 6 interaction layers: home, nursing homes, school, work, university, and community. We incorporated empirical data from multiple datasets available from national, regional, and local sources to infer the connections between the individuals in the different environments introduced by the interaction layers. In Fig.\ref{fig2}, we show the age mixing patterns of the population extracted from our synthetic cities \cite{Mistry2021Jan}.

{\bf Home layer.} Individuals in the home layer are connected if they live together. We extracted the information on the number of homes of a specific size, the average home size, and the home structure at the district level. We use the information from the national census of 2011 \cite{ine2011}, the most recent one that is currently available, (see also the Supplementary Material) for all the cities except for Barcelona, for which this information is available from local sources with higher resolution. We also use the age difference of the home nuclei, at district resolution, from the national census. This information is key for reproducing realistic home contact matrices, as the home structures include in the ``adults'' category any individual aged between 25 and 64 years old. Connecting randomly pairs of individuals in this broad group could lead to less representative links in most homes, composed of two adults alone or with children or old people. As we do not know if these nuclei are assortative or disassortative, we include the data on the age difference of the nuclei to create this synthetic layer.

{\bf School layer. } This layer connects all the students and a teacher within the same scholar unit. Besides, all the teachers that work in the same center are also connected. We included in this layer the infant levels (0-3 and 3-6 years old), primary school (7-12), secondary school (13-16), high school (17-18), and job training (from 17). We inferred these connections using data on the number of students per level, the number of units per level, and the number of schools, taking into account the levels offered by each kind of school. This information was available at the district level for Barcelona and Valencia. Additionally, data on the specific size of each specific unit at each center was available for Valencia and used for that layer inference. For Seville, Zaragoza, and Murcia, this information was available at the municipality level, so we mapped the school coordinates to the districts, and we inferred the rest of the needed information from the one at the municipality level. Once the synthetic units, at each center, were created, we assigned individuals from the population to those units. First, we filled each unit with individuals of that specific age that have their homes located in the same district. Secondly, the units that had not been filled totally with individuals from the same district were filled with individuals from other districts, with a priority determined by the distance between the centroids of the districts, until all the units in the city were full. We assumed that, after that step, the remaining individuals were not included in the education system. Teachers were chosen randomly among individuals aged between 30 and 70 from any district in the city.
 
{\bf University layer.} We generated the university contact layer using the national statistics of the number of registered students per university and per degree, split by sexes, available from the Spanish Ministry of Education, considering both undergraduate and graduate programs, for the academic year 2019-2020. We considered the universities located in the same province as the studied cities, after removing distance-learning universities. Then, we estimated which students are registered and live in the city, and which ones are external, either registered in the same or another province. Local students have also interactions in the other layers, while externals are only in the university layer. Finally, we obtained the national age profiles by sex of university students, picked individuals from our synthetic population, and introduced those external, according to these profiles. We designed a connectivity pattern of all-to-all for degrees that had sizes lower than 50 people, and otherwise generated patches with all-to-all connectivity of a maximum size of 50 people for the larger ones.

{\bf Work layer.} In the work layer, individuals that belong to the same company are connected together. We obtained the size distribution of companies throughout all the Spanish provinces, which follows a power-law distribution pdf($S$)$\sim S^{-2}$. Then, we generated companies with sizes that follow this distribution, and, when sizes were higher than 20 people, we distributed the workers among patches with a maximum size of 20 people. We estimated the number of workers by subtracting the number of autonomous workers from the number of registered workers in each city, according to the Social Security reports. We extracted sex and age features also from Social Security reports on a national scale. We did not consider as potential workers those that were assigned a school patch, either as teachers or students. The synthetic companies were filled with individuals from the synthetic population following the corresponding distributions by age and gender.

{\bf Nursing homes layer.} We collected information on the number of nursing homes and their capacity in each municipality. Additionally, we gathered national statistics on the age and gender of the people that reside in nursing homes. We assumed that the nursing homes need one caretaker for every four places, and chose that uniformly from those in the dataset older than 16 years old (minimum age for being allowed to work). Inside each nursing home, we assumed an all-to-all connection. Note that individuals residing in nursing homes do not interact in the household layer.

{\bf Community layer.} We generated
a synthetic community layer connecting randomly pairs of individuals living in the same district, according to the contact matrices for Spain in Ref. \cite{prem2017projecting}. There were contact matrices available for home, work, school, and other locations, and we chose the latter. This dataset reported the probability of connecting pairs of individuals according to their ages, in age groups of 5 years up to 75 years old. For individuals older than 75 years old, we extrapolated the data of the oldest available group. 

{\bf Spreading model.} We used the COVASIM software for modeling the spread of COVID-19 \cite{kerr2021covasim}. COVASIM is an open-source agent-based modeling tool. This software has been used for studying different scenarios, for example assessing the test-trace-quarantine strategy \cite{kerr2021controlling}, or quantifying the risk of outbreaks after international border opening \cite{pham2021estimating}. We modified COVASIM to include the details of our synthetic cities. Specifically, we included the age, sex, and contacts of the individuals in each of the considered cities. We ran independent simulations where each simulation chose one randomly infected seed as the first infected individual. Then, we kept the endemic realizations, defined as those leading to a finite number of deaths, which we set higher than 10 for the first wave. For the second wave, we also requested that there were more than 10 death observations in the last 10 days of the realization.

{\bf Epidemic data. } We obtained the temporal series of the number of deaths and number of confirmed cases from the Spanish Ministry of Health \cite{cnecovid} at the province level, with daily resolution. Then, we multiplied these values by the fraction of the province's population living in the city. We averaged the rescaled data over a moving window of 7 days (the specific day $\pm$ 3 days) to smooth the fluctuations.
\section{Results}
\subsection{Contact matrices}
The inference of contacts among the synthetic inhabitants of the digital cities that we created facilitated measuring contact matrices, which constitutes fundamental information for informing epidemiological models. The aggregated network of the six contact layers display an assortative pattern where we observe blocks of infants, adults, and the elderly, and within them a higher preference to interact with individuals of similar ages (Fig. \ref{fig2}a-e). The number of contacts per layer is also significantly different both within and across cities (Fig. \ref{fig2}f). For instance, workplace contacts are predominant in Barcelona and Zaragoza, the university ones in Valencia and Murcia, and the school contacts in Seville. 

\subsection{First wave\label{subsec:first}}

\begin{figure*}
	\centering
	\includegraphics[width=\textwidth]{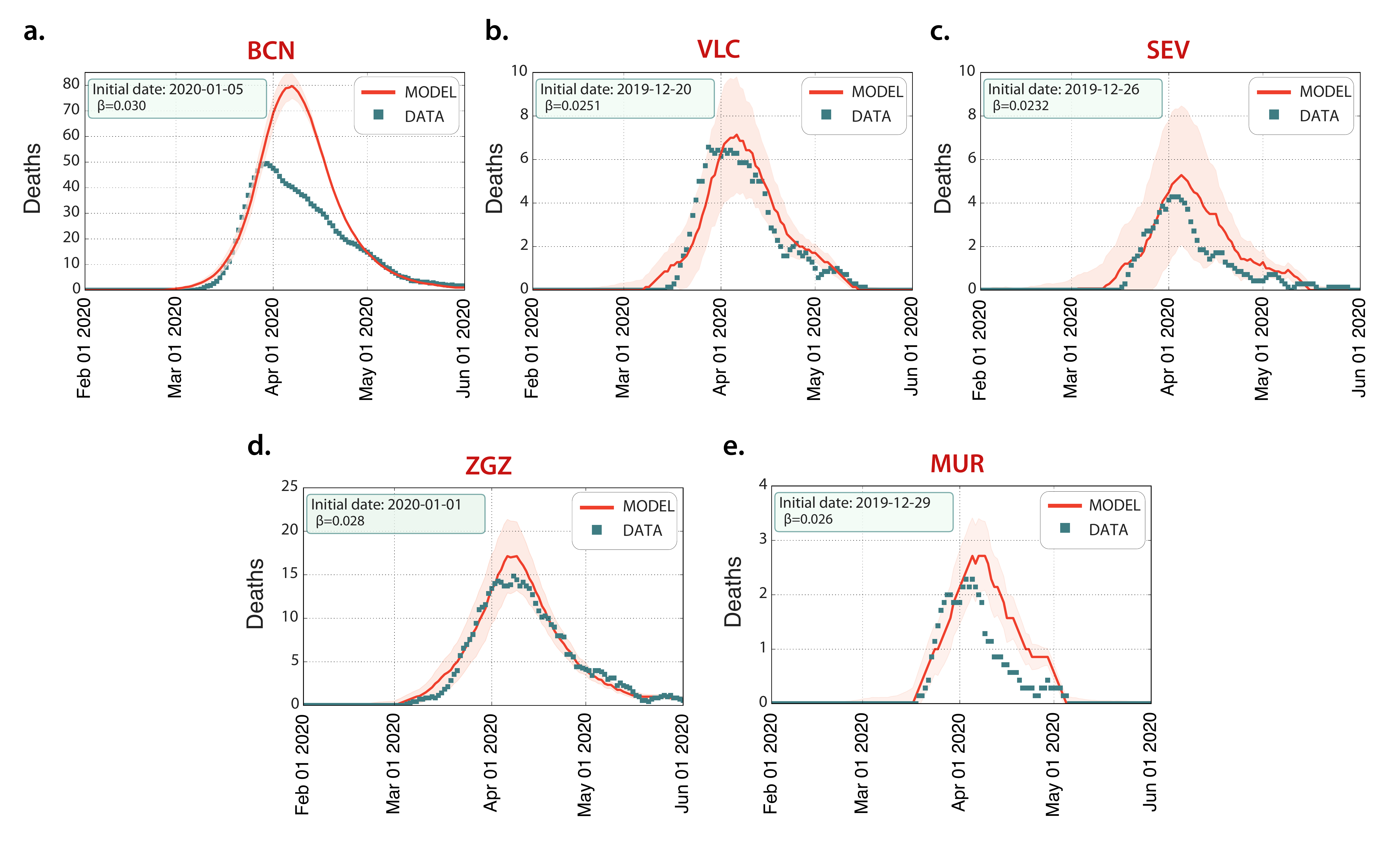}
	\caption{First wave of infections of COVID-19 in five Spanish cities: Barcelona (BCN), Valencia (VLC), Seville (SEV), Zaragoza (ZGZ) and Murcia (MUR). The number of deaths $D$ estimated by our model (solid line, the shaded area represents the 5-95 CI) agrees with the data (properly rescaled, see Methods) reported by the Spanish Ministry of Health (dots), with an excess of deaths in the initial stages of the wave.}
	\label{wave1}
\end{figure*}

In order to be able to explore realistic counterfactuals for the effectiveness of the most important NPIs adopted, we started by simulating the first wave to calibrate the model for each of the cities considered. Specifically, we ran simulations of the spread of COVID-19 in these cities using the software COVASIM. We estimated the transmission rate and the arrival of the initial seed, considered as a single infected individual (Fig. \ref{wave1}, Table \ref{tabwv1}). Our multilayer approach allowed us to introduce the effects of the national lockdown declared on 14th March 2020, reducing the contacts in the work layer to 20\% (10\% for Barcelona) and 0\% in the university, school, and community layers. Our results indicate an earlier arrival of COVID-19 to these cities (upon the assumption of a single initial seed), and they highlight the earlier occurrence of deaths at the beginning of the first wave, not considered in the official statistics, in Barcelona, Valencia, Seville, and Zaragoza. The prevalence estimates from our model are compatible with those obtained from the nationwide seroprevalence study in Spain (see Table \ref{tabwv1}). Note that this seroprevalence study detected 10 times more cases than the ones reported by the surveillance system.

\begin{table*}
	\begin{tabular}{|c|c|c|c|c|c|c|}
		\hline
		{\bf City} & {\bf Start} & $\beta$ & {\bf Prev (5-95 CI)} &  {\bf Deaths (5-95 CI)} & {\bf Ref. Prev. (5-95 CI)} \\ \hline
		Barcelona & Jan 05 & 0.030 & 14.6 (13.1-16.1) & 2334 (2188-2477) & 7.4 (6.2-8.9)\\ \hline
		Valencia & Dec 20 & 0.0251 & 3.1 (1.5-4.7) & 227 (141-313) & 2.1 (1.5-3.0) \\ \hline
		Seville & Dec 26 & 0.0232 & 2.4 (1.1-4.7) & 195 (75-314) & 2.7 (1.9-3.8) \\ \hline
		Zaragoza & Jan 1 & 0.028 & 6.1 (3.9-8.3) & 599 (465-733) & 5.2 (3.9-6.9) \\ \hline
		Murcia & Dec 29 & 0.026 & 2.4 (1.6-3.3) & 89 (67-111) & 1.6 (1.0-2.5) \\ \hline
	\end{tabular}
	\caption{\label{tabwv1}First wave model results. The reference prevalence is that provided in the first phase of the national study of seroprevalence for the first wave in its second round, finished on June 1st \cite{pollan2020prevalence}.}
\end{table*}
\subsection{Second wave. Counterfactuals}
\begin{figure}
	\centering
	\includegraphics[width=0.4\textwidth]{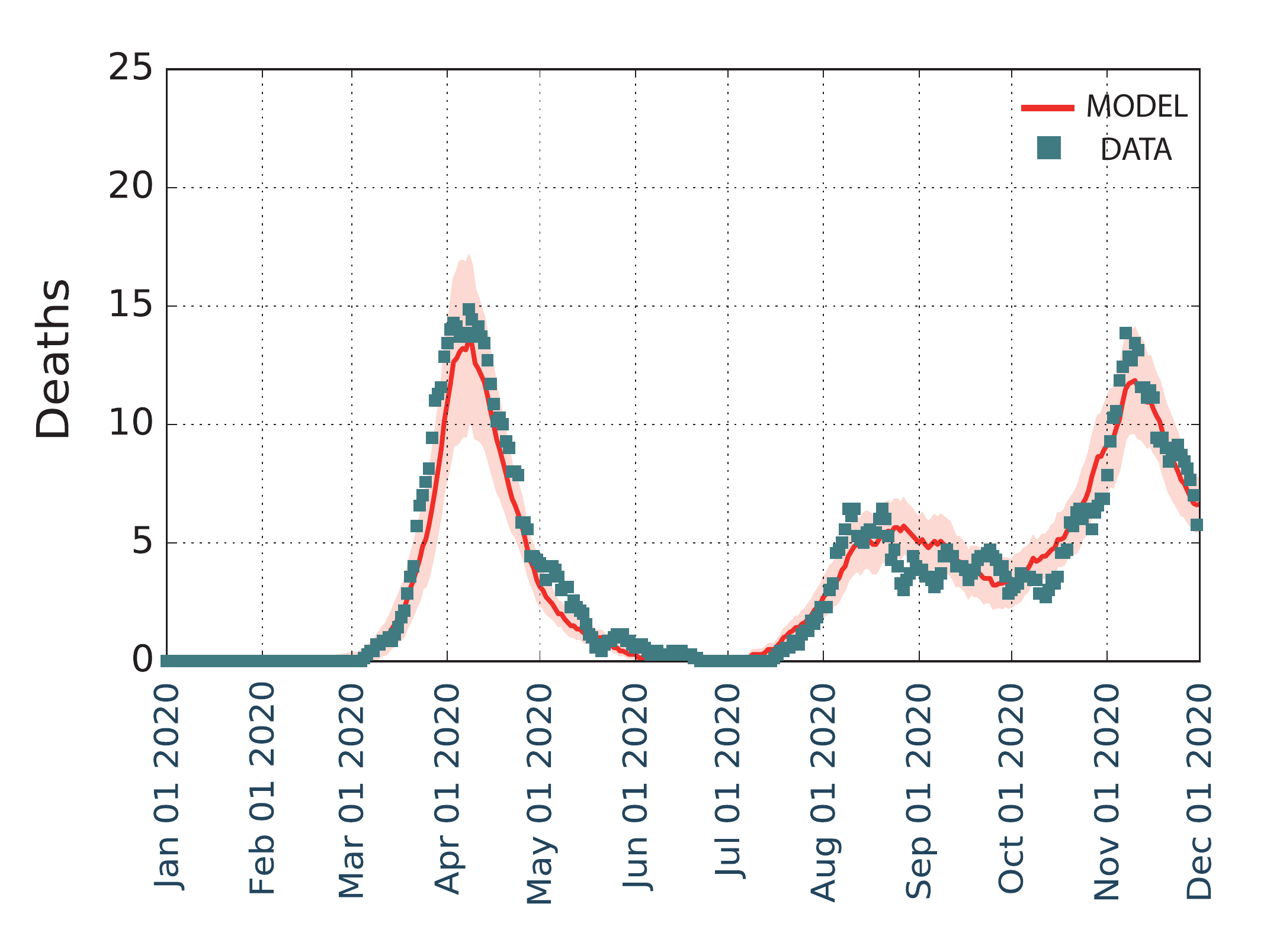}
	\caption{Modeling the first and second waves of COVID-19 spreading on Zaragoza from January to December 2020. We represent the temporal evolution of the number of deaths $D$, with shaded regions depicting the 5-95 confidence intervals of the model.}
	\label{wave2}
\end{figure}

To illustrate the potential of our approach, once we have shown that our model with synthetic digital cities could reproduce the main features of the first wave of the spreading of COVID-19, we implemented a case study of the second wave focusing on the city of Zaragoza. This second wave occurred between July and December 2020.

First of all, let us summarize the non-pharmaceutical interventions that were introduced in this city to mitigate the spread of COVID-19. The main interventions were:
\begin{itemize}
	\item Testing and tracing. Positive tested individuals and their close contacts were isolated for 14 days until 30th September, and from then on for 10 days.
	\item Restrictions on restaurants, cafes, and nightlife. Starting on 5th August, lifted on 4th September, and re-started on 19th October.
	\item Opening of the schools with reduced group sizes and safety measures. Different levels started progressively, from 7th September to 17th September. 
	\item Opening of the universities with reduced group sizes. The university was opened on 14th September.
	\item Interventions impacting the community layer. We considered the interventions related to the State of Alarm and those related to the capacity and schedules of restaurants and bars.
\end{itemize}
In addition to these policies, we also took into account the annual leave period of workers and considered a reduced number of interactions in the work layer starting on 15th July until the middle of September, with the maximum reduction on 1st August. In terms of the model, this implies a higher amount of time (e.g., more weight of this layer) associated with the interactions in the community layer.

We used the same parameters obtained in the calibration of the first wave (i.e., transmission rate $\beta$ and date of arrival of the infected seed), and run the model from the estimated arrival of the first case (see \ref{subsec:first}) to 1st December 2020. Since we ran the simulations from the beginning of the first wave, we lifted progressively the restrictions that were active from the 14th of March in two steps: on the 1st and 20th of June, following the progressive lift of the restrictions that actually happened. Then, a second wave started growing (Fig. \ref{wave2}), and we calibrated our model to obtain the impact of the interventions on our model parameters. 
  
\begin{figure*}
		\centering
	\includegraphics[width=\textwidth]{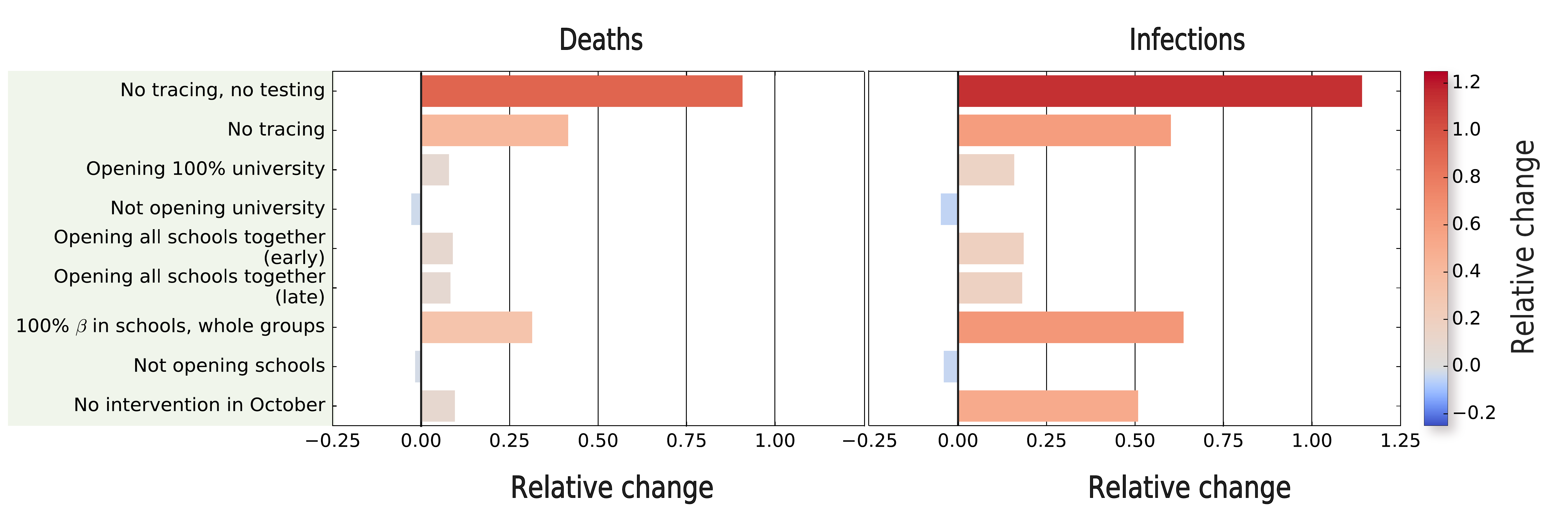}
	\caption{Estimating the impact of non-pharmaceutical interventions on the second wave of COVID-19 spreading in Zaragoza. We compute the relative change with respect to the outcome of the model in terms of prevalence and number of deaths for different simulated scenarios. The relative difference in the relevant quantity $X$ ($X$= deaths or prevalence) is computed as $(X^{\text{counterfactual}}-X^{\text{simulated-2nd-wave}})/X^{\text{simulated-2nd-wave}}$.}
	\label{counterf}
\end{figure*}

We kept 25\% of the contacts in the school and university layers to simulate the small group's policy, and reduced $\beta$ by 50\% in the school layer, considering the strict protocol to avoid contagion at schools. The calibration of the second wave led to the following results:
\begin{itemize}
	\item Varying number of links in the community layer: connections were set at 50\% of the baseline value (1st June, progressive lift of restrictions), 80\% (20th June, end of the national State of Alarm), 50\% (5th August, regional limits on restaurant and bar schedules), 100\% (4th September, lift of restrictions), 30\% (19th October, regional limits on the schedule and capacity of restaurants and bars), and 10\% (26th October, national State of Alarm).
	\item Varying the weight of the links in the community layer: increase by 50\% from 20th June to 26th October (end and beginning, respectively, of the national State of Alarm).
	\item Varying the number of links in the work layer: connections were set at 70\% of the baseline value (1st June), 50\% (15th July), 30\% (1st August), and 50\% (20th September). These changes in summer accounted for the summer holidays period. Mobility reports \cite{google2021} showed a slow return to mobility associated with work in September.
	\item Testing. The probability of symptomatic individuals being tested (per day with symptoms) was estimated to be 15\% between 1st July and 14th September, and 9\% from 15th September. The delay between the test and the result notification (with the beginning of the associated isolation period) was fixed to 1 day.
	\item Contact tracing. The contacts from positive-tested individuals were traced with a general probability $P_t$. Additionally, $P_t$ was weighted for each layer of contacts, fixing the weights 1 for home, 0.8 for school, 0.6 for university, 0.8 for work, 0.0 for the community (until the introduction of the contact tracing app Radar COVID \cite{rodriguez2021population}, which increased it to 0.05), and 0.0 for nursing homes. We fixed the time between the positive notification and the communication with close contacts to 2 days. $P_t$ was estimated to be 0.4 between 1st July and 19th August, 0.45 between 20th August (introduction of Radar COVID) and 30th September, and 0.5 from 1st October (extra support to contact tracing from trained soldiers). 
\end{itemize} 

Overall, the model (Fig. \ref{wave2}) estimates that there were 1354 deaths (5-95 CI), with a prevalence of 22.6\% (21.0\%-24.2\% 5-95 CI). This prevalence is higher than the reported in the fourth phase of the national seroprevalence study carried out in mid November\cite{pollan2020prevalence}, which estimated a prevalence of 12.7\%(10.1-15.8 5-95 CI) at the province level. Even though the data at the municipality level is not available, it reports that the prevalence in municipalities with more than 100,000 inhabitants was 50\% larger than in the smaller ones. The province of Zaragoza is highly heterogeneous in terms of size, with one municipality (out of 293) containing 69\% of the almost 1,000,000 inhabitants in the province. Thus, it is expected that the prevalence at the city level should be much larger. Similarly, our model in the first wave agreed with the empirical observations of the temporal evolution of the number of deaths documented, with a minor overestimation for Murcia but a larger one for Barcelona. We interpret these divergences as possibly missing data, in line with other studies that have claimed a higher number of deaths than that reported by the official statistics, which was particularly significant in the administrative region of Catalonia, where Barcelona is located\cite{garcia2021retrospective}. 

With the model calibrated and the availability of the results of the simulation of the second wave, the next step is to produce different counterfactual scenarios in which we inspect the effect of each of the non-pharmaceutical interventions adopted. More specifically, we computed the number of deaths and the disease prevalence by performing simulations with the same epidemiological parameters but switching on and off alternative interventions. Finally, we computed the relative change in the relevant quantity $X$ ($X$= deaths or prevalence) as $r=(X^{\text{counterfactual}}-X^{\text{simulated-2nd-wave}})/X^{\text{simulated-2nd-wave}}$. Therefore, the absolute change can be obtained as $X^{\text{counterfactual}}=(1+r)\cdot X^{\text{simulated-2nd-wave}}$.

The results depicted in Fig. \ref{counterf} correspond to the following nine different counterfactuals:
\begin{itemize}
	\item No testing and no contact tracing. The testing intervention was removed. Hence, as contact tracing depends on the results of the testing process, contact tracing was automatically removed.
	\item No contact tracing. To analyze the impact of the contact tracing strategy and decouple it from the testing process, we kept the testing intervention and its parameters but removed contact tracing interventions.
	\item Opening 100\% university. We considered the opening of the university layer with 100\% of the contacts, instead of the 25\% contacts estimated through the small group's intervention.
	\item Not opening university. We simulate a scenario where the university layer remained closed.
	\item Opening all schools together (x2). The school opening was done following a staggered strategy, such that each level started on a different date. We simulate scenarios where all the levels started on the same date, either on the date of the earliest opening (7th September) or the latest opening (17th September).
	\item 100\% $\beta$ in schools, whole groups. Schools were one of the sectors where strong protocols were introduced, reducing considerably the infection rate and also the group size. We simulated the absence of these protocols, keeping the same infection rate as in the rest of the layers, and considering this layer with whole groups, that is, 100\% of the contacts.
	\item Not opening schools. We quantified the changes in the outcome of the second wave if the schools had not been open.
	\item No interventions in October. We observe that the interventions in October were key to controlling the second wave. Thus, we removed these interventions and computed this counterfactual, keeping the same final date, such that the result is comparable with the rest of the counterfactuals. However, we assumed that removing these interventions would imply that the second wave continued growing on time. To characterize this growth (in terms of both time extent and outcome), we ran additional simulations for 15 and 30 days more and compared them with extrapolations of the original second wave simulation for the same period (without including additional measurements introduced on December 2020 or calibrating the observed data in that period). 
\end{itemize}

The results of the counterfactual analysis shown in Fig. \ref{counterf} indicate that the combination of tracing and testing, with the associated isolation of positive individuals, was very effective in reducing the number of both deaths and infections. Note that for this case, the counterfactual (i.e., lack of such measures) produces more than twice the number of infected individuals, and also nearly doubles the number of deaths. Next, we quantified the impact of contact tracing alone by keeping the testing process, together with the isolation of individuals with a positive test, but removing the contact tracing. This scenario also shows an increase in both the number of infections and deaths. However, the increase of both observables is twice lower than if both contact-tracing and testing are removed, highlighting the importance of combining these two interventions to achieve the best result. The third-most important counterfactual according to the increase in deaths was the removal of the interventions in schools, which however produce the second-largest increase in the number of infections, but with a smaller impact in the number of deaths because most infections would occur in non-risk age groups. Interestingly, opening the university without restrictions would lead to fewer infections than with schools completely opened, however leading to more deaths. The synchronous opening of the schools for different levels also implied an increase in prevalence and deaths, but with a lower impact than other interventions. Finally, there are also some counterfactual scenarios that produce a decrease in the observables, such as keeping schools or the university closed. However, their impact is minor.  

For the sake of completeness, we also assessed the impact of the interventions in October to control the outbreak. In principle, these measures did not have a big impact as shown in Fig. \ref{counterf}. However, the interpretation is not straightforward, because our second wave simulations finish on December 1st, and the absence of these interventions may have implied a later end of this wave. To quantify this, we extended, without adding any new intervention, both the calibrated and counterfactual simulations, finishing the simulation on a) 15th December and b) 31st December. Our results showed that the relative change between the real extrapolated framework and this counterfactual kept increasing after 1st December (9.5\% for deaths, 50.9\% for prevalence), as the extrapolated values were higher on 15th December (40.3\% for deaths, 71.1\% for prevalence), and slightly decreased at 31st December (25.3\% for deaths, 66.5\% for prevalence), indicating that by this date, the counterfactual wave would have finished. Hence, this extrapolation suggests that this counterfactual would have implied, compared with the rest of the counterfactuals, the second-largest increase in prevalence, and the fourth-largest increase in the number of deaths.

\section{Discussion and Conclusions.}
The quick spread of a deadly infection among a population represents a threat to public health systems, which requires immediate action and extra resources in order to mitigate and eventually control the impact of the associated disease on the population. However, interventions need to be carefully evaluated, as our society is a complex interdependent system in which mitigating the effects of one disease might result in non-desirable side effects such as the reduction of the services provided to both prevent and treat other diseases \cite{roberts2020pandemic,jewell2020potential,tovar_modeling_2022}. 
To avoid these effects, scientists and policy-makers need computational resources that allow for a fast assessment of the possible outcome of interventions whenever pharmaceutical interventions, such as vaccination campaigns, are not possible. This is generally the case when new diseases emerge, which currently represent one of the major threats to public health systems. This is because, on the one hand, an emerging disease needs the allocation of a significant amount of resources in order to develop and test vaccines and/or specific pharmacological treatments. On the other hand, new diseases, as has been the case of COVID-19, are likely to escape traditional surveillance methods, leading to a high number of undocumented infections during the early stages of the disease, as also happened in the case of SARS-CoV-2 \cite{li2020substantial}. In such situations, non-pharmaceutical interventions represent the alternative to at least earning time. This is where data-driven and computational frameworks are fundamental to inform models that can illustrate the outcome of different non-pharmaceutical interventions. In this paper, we have presented a model that could be used to characterize the consequences of a plethora of non-pharmaceutical interventions. We applied the model to study the first and second waves of COVID-19 in Spain, finding that testing, tracing, and isolation were among the most effective interventions to reduce both the number of deaths and infections, in line with similar studies for other geographical locations \cite{aleta2020modelling,steinbrook2020contact,salathe2020covid}. Our study also shows that, on the whole, the interventions adopted during the second wave for the city of Zaragoza, were effective and reduced the number of deaths and infections by around 10\% and 50\%, respectively. 

This work has some limitations that deserve further discussion. First of all, our simulations considered a single randomly chosen initial seed, and from this, we estimated the date of arrival of the disease. Nevertheless, the spreading process could have started by the arrival of several infected individuals either synchronously or asynchronously. However, we think that these approaches are equivalent, as they would lead to the same number of infected individuals at later dates. In contrast, different effects could emerge when specific individuals, according to, for example, their age, district of residence, or employment status, display a higher likelihood to introduce the infection. Another limitation is the isolation of the cities, as they are considered closed systems. This can be solved by introducing a spontaneous infection rate reflecting the imported cases from other locations, although we assume that in the cases of generalized local transmission this rate would lead to minor differences. The emergence of variants with different infection and recovery rates and death probabilities is challenging for these models, requesting the parameter correction for subsequent waves happening when other variants were present. The latter, however, does not impact predictions at the early stages of an emerging disease, which is when evaluating possible NPIs is most needed. Finally, we considered the main non-pharmaceutical interventions that were applied in the city of Zaragoza, but their calibration may include the effect of other interventions that we assumed to have minor effects. 

In summary, our work shows how models of digital cities can be coupled to agent-based epidemiological models of disease dynamics and be used for scenario evaluation. 

\section*{Conflict of Interest Statement}

The authors declare that the research was conducted in the absence of any commercial or financial relationships that could be construed as a potential conflict of interest.

\section*{Author Contributions}

JPR, AA, and YM contributed to conception and design of the study. JPR collected, cleaned and organized the data. JPR performed the statistical analysis. JPR, AA, and YM analyzed the results. JPR wrote the first draft of the manuscript. All authors contributed to manuscript revision, read, and approved the submitted version.

\section*{Funding}

J.P.R., A.A., and Y.M. acknowledge support from the Government of Aragon (FONDO–COVID19-UZ-164255). J.P.R. is supported by Juan de la Cierva Formacion program (Ref. FJC2019-040622-I) funded by MCIN/AEI/ 10.13039/501100011033. A.A. acknowledges support through the grant RYC2021‐033226‐I funded by MCIN/AEI/10.13039/501100011033 and the European Union ``NextGenerationEU/PRTR''. Y.M was partially supported by the Government of Aragon, Spain and ``ERDF A way of making Europe'' through grant E36‐20R (FENOL), and by Ministerio de Ciencia e Innovación, Agencia Espa\~nola de Investigación (MCIN/AEI/10.13039/501100011033) Grant No. PID2020‐115800GB‐I00. The authors acknowledge the use of the computational resources of COSNET Lab at Institute BIFI, funded by Banco Santander through grant Santander‐UZ 2020/0274 and by the Government of Aragon (FONDO–COVID19-UZ-164255). The funders had no role in the study design, data collection, analysis, decision to publish, or preparation of the manuscript.

\section*{Acknowledgments}
The authors acknowledge the Department of University Statistics of the Spanish Ministry of Universities for facilitating the number of registered students at the university per municipality, which is not currently publicly available online.

\section*{Data Availability Statement}
The data sources are specified in the Supplementary Material. Digital Cities and the code are available from the authors upon reasonable request.

\bibliography{manuscript}

\end{document}